\documentclass{pasj00}

\begin{document}
\SetRunningHead{Sugawara et al.}{Suzaku Observation of Abell 2319}
\Received{2009/06/22}
\Accepted{2009/09/07}

\title{Suzaku Observation of the Radio Halo Cluster Abell 2319: 
       Gas Dynamics and Hard X-ray Properties}


\author{%
   Chika \textsc{Sugawara}\altaffilmark{1}
   Motokazu \textsc{Takizawa}\altaffilmark{2}
   and
   Kazuhiro \textsc{Nakazawa}\altaffilmark{3}
       }
 \altaffiltext{1}{Graduate School of Science and Engineering, Yamagata University,
                  Kojirakawa-machi 1-4-12, Yamagata 990-8560}
 \altaffiltext{2}{Department of Physics, Yamagata University,
                  Kojirakawa-machi 1-4-12, Yamagata 990-8560}
 \email{takizawa@sci.kj.yamagata-u.ac.jp}
 \altaffiltext{3}{Department of Physics, The University of Tokyo, 
                  7-3-1 Hongo, Bunkyo-ku, Tokyo 113-0033}
%


\maketitle

\begin{abstract}
We present the results of Suzaku observation of the radio halo cluster Abell 2319.
The metal abundance in the central cool region is found to be higher than the surrounding region,
which was not resolved in the former studies. 
We confirm that the line-of-sight velocities of the intracluster medium 
in the observed region are consistent with those of the member galaxies of entire A2319 and 
A2319A subgroup for the first time, though any velocity difference
within the region is not detected. On the other hand, we do not find any signs
of gas motion relevant to A2319B subgroup.
Hard X-ray emission from the cluster is clearly detected, 
but its spectrum is likely thermal. Assuming a simple single temperature model for the
thermal component, we find that the upper limit of the non-thermal inverse Compton component becomes
$2.6 \times 10^{-11}$ erg s$^{-1}$ cm$^{-2}$ in the 10-40 keV band, 
which means that the lower limit of the magnetic field is 0.19 $\mu$G
with the radio spectral index 0.92. 
Although the results slightly depend on the detailed spectral modeling,
it is robust that the upper limit of the power-law component flux and lower limit of the magnetic field
strength become $\sim 3 \times 10^{-11}$ erg s$^{-1}$ cm$^{-2}$ and $\sim 0.2 \mu$G, respectively.
Considering the lack of a significant amount of very hot ($\sim 20$ keV) gas and 
the strong bulk flow motion, it is more likely that the relativistic non-thermal electrons 
responsible for the radio halo
are accelerated through the intracluster turbulence rather than the shocks.
\end{abstract}

\section{Introduction}
Diffuse non-thermal synchrotron radio emission is found in a significant fraction of 
galaxy clusters \citep{Giov00, Kemp01}, 
which indicates that there exist both the relativistic electrons and 
magnetic field as well as the thermal intracluster medium (ICM)
in the intracluster space. Some of such diffuse radio sources
are located near the center and cover the cluster entirely, 
which are called ``radio halos''. Others are located at periphery and called
``radio relics''. 

Although the origin of these non-thermal electrons is still unclear, 
some connections of radio halos and relics with dynamical motion of ICM
are reported. For example, they are rarely found in cool core clusters
\citep{Giov99}.
Radio luminosity has strong correlation with X-ray one and ICM temperature
\citep{Lian00}. 
In addition, \citet{Buot01} shows the correlation of the radio power with asymmetry of 
the X-ray image.

It is likely that such non-thermal electrons are accelerated via shocks
\citep{Sara99, TakiNait00, ToKi00, Mini01, Ryu03, Taki03, Inou05} 
and/or turbulence \citep{Rola81, Schl87, Blas00, Ohno02, Fuji03, Brun04}
in the ICM associated with cluster mergers. 
Magnetic field plays a crucial role in these acceleration processes. However, 
it is not so easy to determine cluster magnetic field strength and structures
in an observational way.
Faraday rotation measure observations of polarized radio sources in and/or behind
clusters are often used for this purpose \citep{Clar01, Vogt03, Govo06}. 
However, the results strongly depend
on the magnetic field structures themselves. Moreover, it can be applied for very
limited regions where we have suitable polarized radio sources by chance, though
this will be possibly resolved by using cosmic microwave background (CMB) as
the polarized sources \citep{Ohno03, Taki08}. 
Therefore, it is very important to estimate the magnetic field strength
through another independent method.

GeV electrons in radio halos and relics are expected to emit non-thermal 
hard X-ray via inverse Compton process of CMB photons. Comparing the synchrotron
radio flux and inverse Compton hard X-ray one (or its upper limit), we are able to estimate
volume-averaged magnetic field strength (or its lower limit).
However, it is very difficult to detect such components 
because of a large amount of thermal emission from the ICM.
Although a lot of efforts have been made to detect the non-thermal inverse Compton 
component from clusters of galaxies, the situation is still unclear.
For example, detection of non-thermal hard X-ray from the Coma cluster 
were reported from the Beppo-SAX PDS \citep{Fusc99, Fusc05} and RXTE \citep{Reph99, Reph02}
data, but their reliability is still controversial \citep{Ross04, Fusc07}. 
Recently, in addition, \citet{Wik09} show
the lower limit that conflicts with the former detection reports through combined analysis
of Suzaku and XMM. Therefore, firm detection or 
reliable upper limits of non-thermal components in the hard X-ray
regime are highly desired by independent instruments.

Abell 2319 is one of the most well-known examples of merging clusters with a giant 
radio halo.
Two subgroups, A2319A and A2319B, are recognized in radial velocity distribution of 
the member galaxies, which suggests that the merger axis is nearly along 
the line-of-sight and that the velocity difference between them is almost
3000 km s$^{-1}$ \citep{Oege95}.
\citet{Mark96} studies temperature structures in a relatively large scale and finds that
the temperature in the region corresponding to the A2319B is lower.
Chandra observations reveal inhomogeneous temperature 
structures and a cold front in the central region, and show that
the position of the X-ray center is different than that of the cD galaxy 
\citep{Ohar04, Govo04}.
Thus, it is obvious that the cluster is not dynamically relaxed.
A2319 has a giant radio halo, and the radio and thermal X-ray distributions are quite
similar to each other \citep{Fere97, Govo01}.
Observations in the hard X-ray band are performed by Beppo-SAX \citep{Mole99},
RXTE \citep{Grub02}, and Swift-BAT \citep{Ajel09}, any of which do not 
report firm detection of non-thermal components.

The hard X-ray detector (HXD) on board Suzaku 
\citep{Mits07} is superior in order to investigate hard X-ray properties of
galaxy clusters because of its low detector background and narrow field of view
\citep{Koku07, Taka07}.
Indeed, improved constraint for non-thermal components has been obtained
for several clusters \citep{Fuji08, Kawa09, Naka09, Wik09}, and very hot
components ($\sim$ 20 keV) were found in RXJ1347.5-1145 \citep{Ota08} 
and A3667 \citep{Naka09}.
In addition,
high spectral resolution of the XIS \citep{Koya07} on board Suzaku 
makes it possible to constrain the bulk flow motion of the ICM that is likely
relevant to the particle acceleration processes \citep{Ota07, Fuji08}.

In this paper, we present Suzaku observation of the Abell 2319 cluster 
to investigate dynamical status of the ICM and hard X-ray properties,
which enables us to understand particle acceleration processes and 
the origin of the radio halo.
The canonical cosmological parameters of $H_0=70$ Mpc$^{-1}$ km s$^{-1}$, 
$\Omega_0=0.3$ and $\lambda_0=0.7$ are used in this paper. Because the cluster
mean redshift is $z=0.0557$, 1' corresponds to 62 kpc.
Unless otherwise stated, all uncertainties are given at the 90\% confidence level.

The rest of this paper is organized as follows. In \S 2 we describe 
the observation and data reduction.
In \S 3 we present the spectral analysis results. In \S 4 we discuss the results
and their implications.
In \S 5 we summarize the results.

\section{Observation and Data Reduction}
\begin{figure}
  \begin{center}
    \FigureFile(80mm,80mm){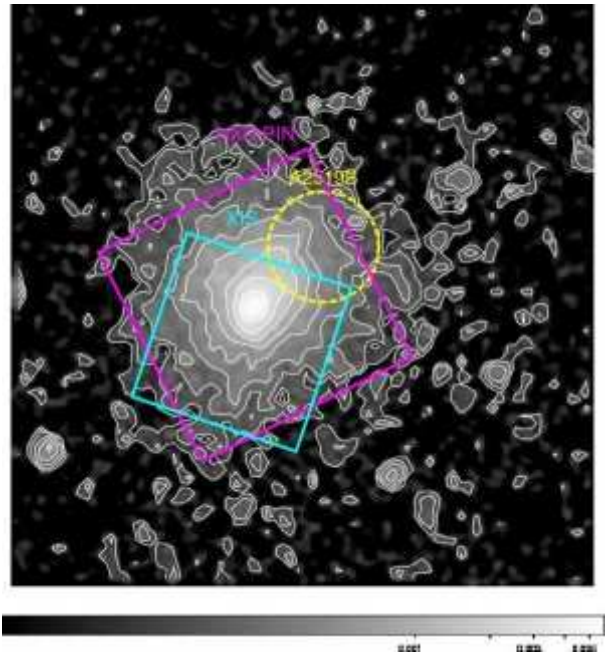}
  \end{center}
  \caption{A ROSAT PSPC image of the Abell 2319 overlaid with field of view of 
           Suzaku XIS CCDs (blue), and that of HXD PIN (magenta) 
           in which the throughput of the fine-collimator becomes 50\%.
           An approximate position of the subgroup A2319B is also represented by dotted yellow circle.
           }\label{fig1}
\end{figure}

We observed the central region of Abell 2319 with Suzaku on 2006 October 27-30 for an
exposure time of 100 ks. The field of view (FOV) of Suzaku XIS, and that of HXD PIN
in which the throughput of the fine-collimator becomes 50 \%  are shown in 
a ROSAT PSPC image (figure \ref{fig1}). An approximate position of the A2319B subgroup is 
also shown. The observation was performed at HXD nominal pointing.
As a result, most of the A2319B is not within the XIS FOV. The XIS was operated 
in the normal full-frame clocking mode. The edit mode was 3 $\times$ 3 and 5 $\times$ 5,
and we used the combined data of both modes. A half of the 64 PINs of the HXD were
operated in the nominal bias voltage of 500 V, and the others were done in that of 400 V.
All the data was processed with the Suzaku pipeline processing of version 2.0.6.13.
We employed the calibration data files of 20081009.

\begin{figure}
  \begin{center}
    \FigureFile(80mm,80mm){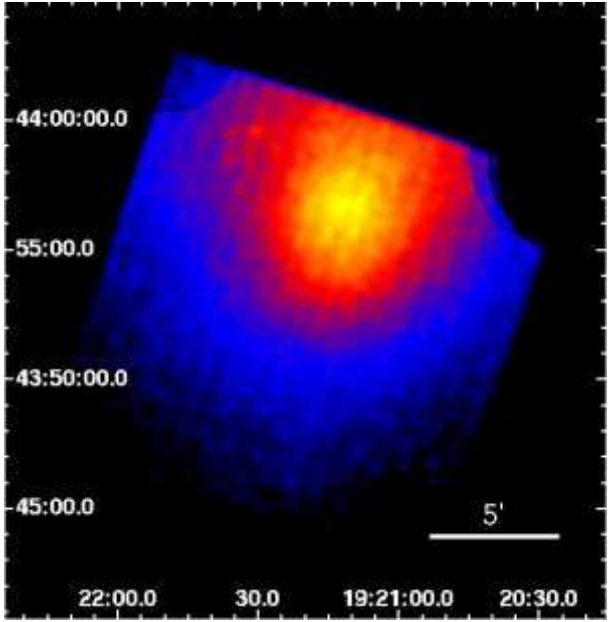}
  \end{center}
  \caption{XIS image of the A2319 cluster in 0.5-8.0 keV band combined from the FI CCD
           images. The images were corrected for exposure and vignetting effects 
           after subtracting NXB, and smoothed by a Gaussian kernel with $\sigma=0.3$ arcmin.
           }\label{fig2}
\end{figure}

The XIS data were processed through standard criteria as follows. Events with a GRADE
of 0,2,3,4,6 and STATUS with 0:524287 were extracted. We exclude the data obtained at
the South Atlantic Anomaly (SAA), within 436 sec after the passage of SAA, and at the low
elevation angles from the Earth rim of $< 5^{\circ}$ and the sun-lit Earth rim of $< 20^{\circ}$.
As a result, effective exposure time becomes 99.5 ks for XIS.
Non X-ray background (NXB) spectra and images of XIS were generated using the ftool
``xisnxbgen'' \citep{Tawa08}.
Figure \ref{fig2} represents the 0.5-8.0 keV XIS image combined from those of 
the front illuminated (FI) CCDs (XIS0, XIS2, XIS3).
The images were corrected for exposure and vignetting effects after subtracting NXB,
and smoothed by a Gaussian kernel with $\sigma=0.3$ arcmin.

The HXD data were also processed in a standard way. We exclude the data obtained at
SAA, within 500 sec after the passage of SAA, within 180 sec before entering SAA,
at the low elevation angles from the Earth rim of $< 5^{\circ}$, and at the location 
where the Cut-Off-Rigidity (COR) is lower than 6 GV. The resultant effective exposure time 
is 94.0 ks. As the NXB of the PIN, we used 
a public NXB model provided by the HXD team in the Suzaku website. The version of the model
is ``METHOD=LCFITDT'', or ``tuned'' \citep{Fuka09}.
The CXB level is estimated in the same way as \citet{Naka09} from the Lockman hole
observation (Suzaku observation ID, 101002010). The detailed procedure is described
in appendix 2 of \citet{Naka09}. We defined the photon flux model as 
$N(E)=8.7 \times 10^{-4} \times E^{-1.29} \times \exp(-E/40.0)$ in photons cm$^{-2}$
s$^{-1}$ keV$^{-1}$ FOV$^{-1}$, where $E$ is the photon energy in keV.

\section{Spectral Analysis}

For spectral analysis of the XIS data, rmf and arf files are generated using the ftool 
``xisrmfgen'' and ``xissimarfgen'', respectively. The ROSAT PSPC image of the A2319 is used
to make arf files, considering that the XIS FOV does not cover the cluster ICM emission entirely
and that Suzaku has only moderate spatial resolution.
The cluster ICM emission in the XIS FOV is so bright that the background components from 
astrophysical origin such as CXB and Galactic diffuse components are safely negligible. 

We use an rmf file provided by the HXD team in the CALDB for PIN spectral analysis.
A PIN arf file are made in the same way of \citet{Naka09}; we calculate it 
by convolving the point source arf assuming that the spatial distribution of the emission is
the same as that of the ROSAT PSPC image. Detailed description is written in 
subsection 2.4 of \citet{Naka09}.

\subsection{Temperature and Abundance Structures}
\label{ss:tem_ab}

\begin{figure}
  \begin{center}
    \FigureFile(80mm,80mm){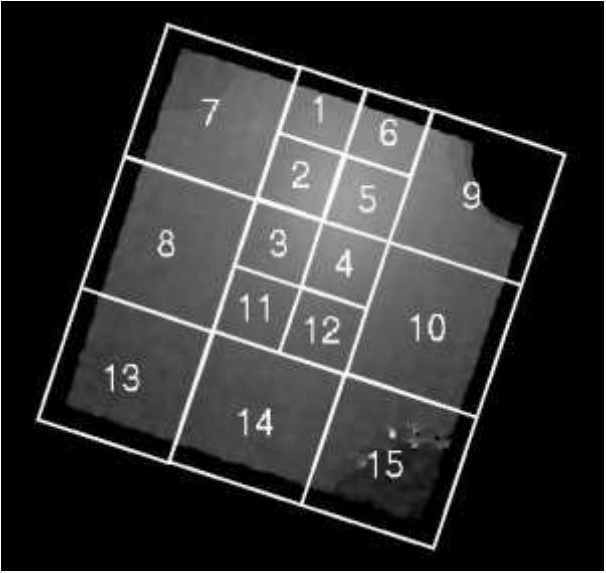}
  \end{center}
  \caption{Region numbers used in two-dimensional 
           temperature and abundance distribution measurement 
           in figure \ref{fig4} and \ref{fig5}.
           }\label{fig3}
\end{figure}

\begin{figure}
  \begin{center}
    \FigureFile(80mm,80mm){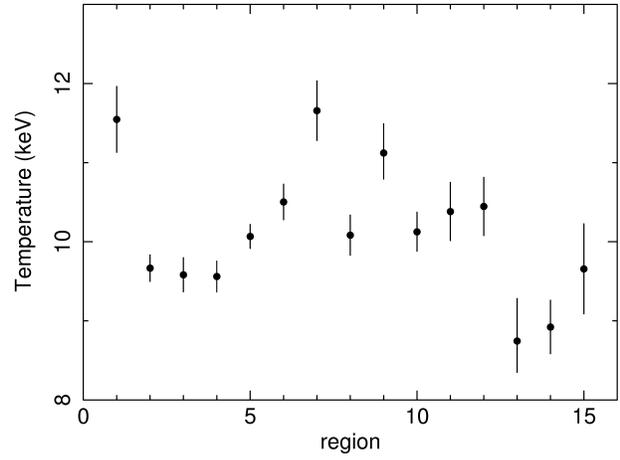}
  \end{center}
  \caption{Temperature for the regions presented in figure \ref{fig3}.
           }\label{fig4}
\end{figure}

\begin{figure}
  \begin{center}
    \FigureFile(80mm,80mm){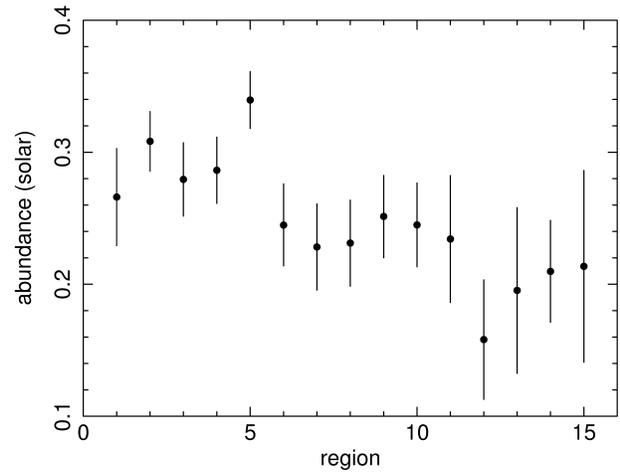}
  \end{center}
  \caption{Same as figure \ref{fig4}, but for the metal abundance.
           }\label{fig5}
\end{figure}

Temperature and abundance structures can provide us with basic and crucial information about 
the ICM dynamics and its origin. 
To explore this, we divide the region observed with XIS into 15 regions 
presented in figure \ref{fig3} and perform spectral analysis for each region. The sizes of
the divided region are 3' $\times$ 3' and 6' $\times$ 6' for the cluster center and surrounding
regions, respectively. FI CCD spectra and related response files for each region are summed,
and FI and back illuminated (BI) CCD spectra are fitted simultaneously. 
For the spectral fitting, we use the energy band of
0.6 -- 10.0 keV and 0.6 -- 8.0 keV for XIS FI and BI, respectively, though the energy band near
the Si edge (1.7 -- 1.9 keV) is ignored to avoid uncalibrated structures. Each spectrum is fitted
by the photoabsorbed single temperature APEC model (WABS $\times$ APEC), assuming that 
$N_H=7.93 \times 10^{20}$ cm$^{-2}$ \citep{Dick90} and that redshift=0.0557 \citep{Stru99}.
The fitting results are presented in table \ref{tab:kt_abun}.
In general, each data are fitted well with the above-mentioned model. Figure \ref{fig4} and
\ref{fig5} show the resultant temperature and metal abundance for the regions presented in
figure \ref{fig3}, respectively. The central south-east regions (region 2, 3, and 4)
tend to have lower temperature and higher abundance.
\begin{table}
  \caption{Best fit parameters for the XIS spectra of the regions presented in figure 
           \ref{fig3}.}
  \label{tab:kt_abun}
  \begin{center}
    \begin{tabular}{llll}
      \hline\hline
      region   &  $kT$  &  $Z$       &  $\chi ^2$/d.o.f    \\
               &  (keV) &($Z_{\odot}$)&                     \\
      \hline
      1   &  $11.5^{+0.4}_{-0.4}$  &   $0.27^{+0.04}_{-0.04}$  & 977.3/892 \\
      2   &  $9.7^{+0.2}_{-0.2}$   &   $0.31^{+0.02}_{-0.02}$  & 1540.0/1389 \\
      3   &  $9.6^{+0.2}_{-0.2}$   &   $0.28^{+0.03}_{-0.03}$  & 967.9/982  \\
      4   &  $9.6^{+0.2}_{-0.2}$   &   $0.29^{+0.03}_{-0.03}$  & 1198.3/1132 \\
      5   &  $10.1^{+0.2}_{-0.2}$  &   $0.34^{+0.02}_{-0.02}$  & 1775.7/1610 \\
      6   &  $10.5^{+0.2}_{-0.2}$  &   $0.24^{+0.03}_{-0.03}$  & 1117.3/990  \\
      7   &  $11.7^{+0.4}_{-0.4}$  &   $0.23^{+0.03}_{-0.03}$  & 1162.3/1079 \\
      8   &  $10.1^{+0.3}_{-0.3}$  &   $0.23^{+0.03}_{-0.03}$  & 953.6/902   \\
      9   &  $11.1^{+0.4}_{-0.3}$  &   $0.25^{+0.03}_{-0.03}$  & 1219.0/1127 \\
      10  &  $10.1^{+0.2}_{-0.2}$  &   $0.24^{+0.03}_{-0.03}$  & 1012.3/934  \\
      11  &  $10.4^{+0.4}_{-0.4}$  &   $0.23^{+0.05}_{-0.05}$  & 501.3/449 \\
      12  &  $10.4^{+0.4}_{-0.4}$  &   $0.16^{+0.05}_{-0.05}$  & 420.2/445 \\
      13  &  $8.7^{+0.5}_{-0.4}$   &   $0.20^{+0.06}_{-0.06}$  & 283.0/282 \\
      14  &  $8.9^{+0.3}_{-0.3}$   &   $0.21^{+0.04}_{-0.04}$  & 623.1/551  \\
      15  &  $9.7^{+0.6}_{-0.6}$   &   $0.21^{+0.07}_{-0.07}$  & 362.8/251  \\
      \hline
    \end{tabular}
  \end{center}
\end{table}

\begin{figure}
  \begin{center}
    \FigureFile(80mm,80mm){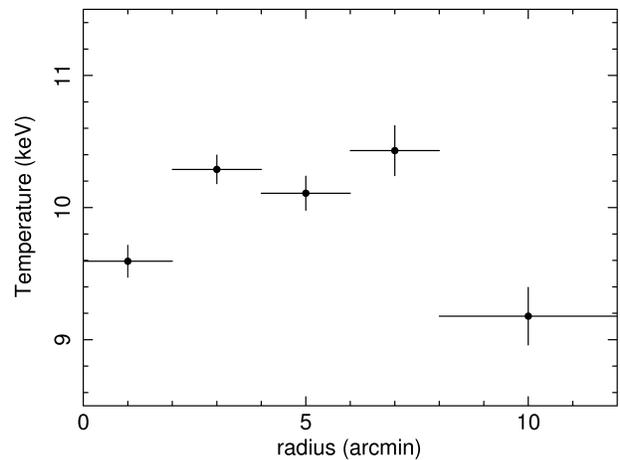}
  \end{center}
  \caption{Projected radial temperature profile.
           }\label{fig6}
\end{figure}

\begin{figure}
  \begin{center}
    \FigureFile(80mm,80mm){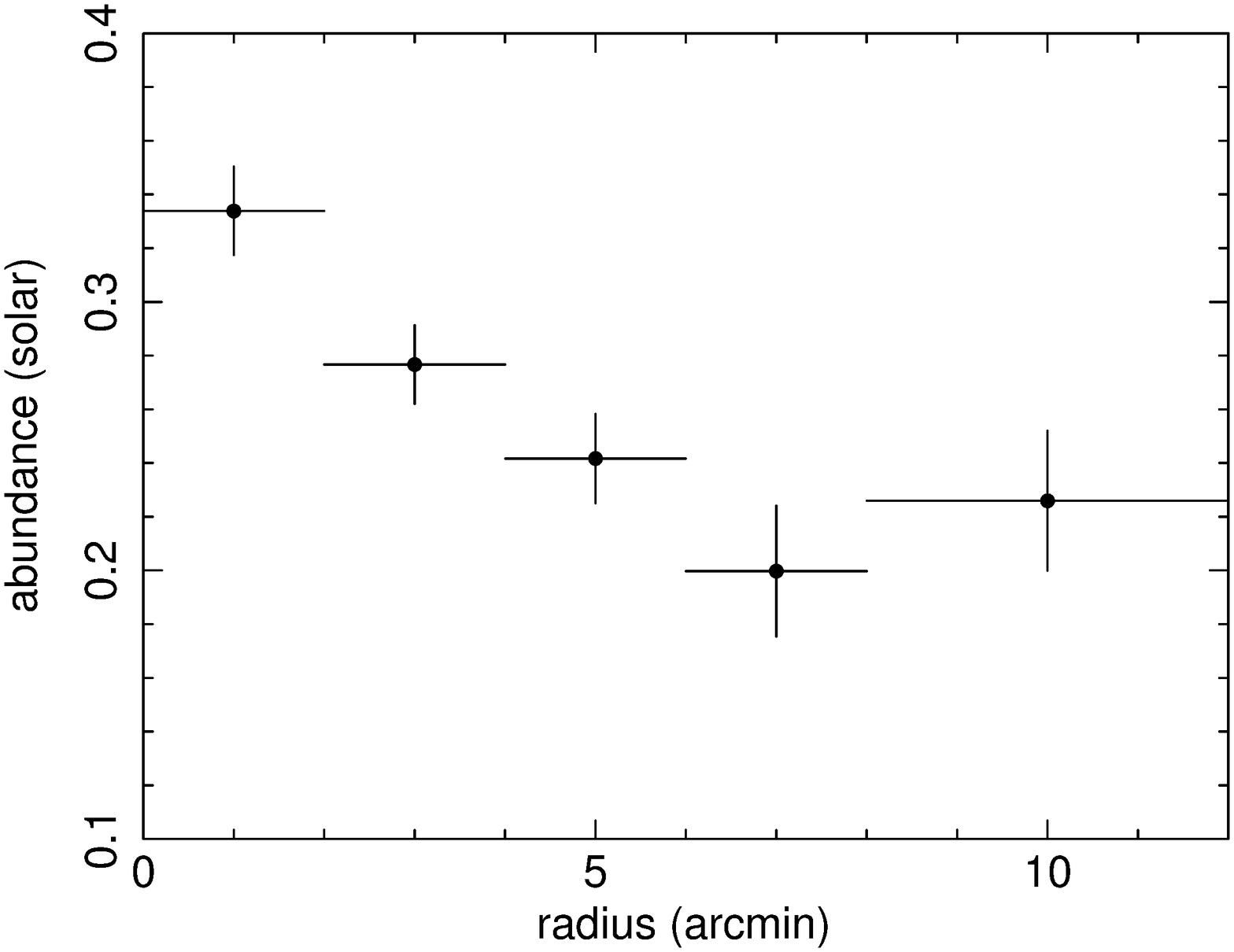}
  \end{center}
  \caption{Same as figure \ref{fig6}, but for the metal abundance.
           }\label{fig7}
\end{figure}

In order to confirm this trend of temperature and metal abundance distribution, we determined
the X-ray spectra in annuli
whose center is located at the X-ray peak $(290.3000, +43.9458)$. 
The details of the spectral fitting procedure are similar 
in 2D analysis presented above. 
Figure \ref{fig6} and \ref{fig7} show the radial (not deprojected) 
profile of the temperature and abundance, respectively. 
It is clear that the ICM in the central region has lower temperature and higher abundance.

\subsection{Line-of-sight Velocities of the ICM}
\begin{figure}
  \begin{center}
    \FigureFile(80mm,80mm){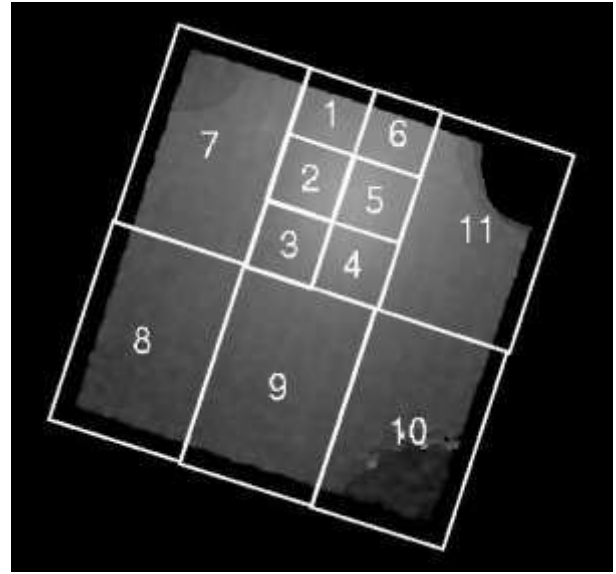}
  \end{center}
  \caption{Region numbers used in measurement of the line-of-sight velocity of the ICM
           in figure \ref{fig9}.
           }\label{fig8}
\end{figure}

Line-of-sight velocities of the ICM contain more direct information about the cluster 
dynamics than X-ray surface brightness and temperature structures. 
In general, however, more photons are necessary to determine the center of the Doppler-shifted lines 
with meaningful accuracy through the spectral fit of the XIS data than to determine
the temperature and metal abundance, which means that we have to use larger regions.
Thus, we divide XIS FOV into 11 regions shown in figure \ref{fig8} and perform spectral analysis
for each region. The sizes of a region are 3' $\times$ 3' and 6' $\times$ 9' for the 
central regions (from 1 to 6) and the outer parts (from 7 to 11), respectively.
In order to determine ICM line-of-sight velocities, we mainly use Doppler-shifted
He like Fe K$\alpha$ lines (6.679 keV) and H like Fe K$\alpha$ line (6.964 keV). 
Therefore, if the ICM motion is $\sim$ 1000 km s$^{-1}$, 
we should resolve the energy shift of only $\sim$ 22 eV. This is a challenging task considering 
that energy resolution of XIS is 130 eV (FWHM). 

To measure an energy scale of XIS accurately, we make a gain correction in a similar way of 
\citet{Fuji08} with Mn K$\alpha$ lines of the calibration sources on each XIS sensor. We made spectra 
of the calibration source region for each XIS sensor, and fit
them with APEC plus two Gaussians assuming that the abundance and redshift of APEC are equal
zero.  A gain correction factor is defined as
\begin{eqnarray}
  f_{\rm gain} = \frac{E_{\rm fit}({\rm Mn-K}\alpha)}{E_0({\rm Mn-K}\alpha)},
\end{eqnarray}
where $E_{\rm fit}$(Mn-K$\alpha$) and $E_0$(Mn-K$\alpha$) are the
energy value obtained from the fitting results and expected one (5.895 keV),
respectively. We evaluate a gain correction factor $f_{\rm gain}$ for
each XIS sensor. Typically, we find $|f_{\rm gain}-1|$ to be $\sim 0.0005$, which
corresponds to be $\sim 150$ km s$^{-1}$.
Then, the corrected redshift becomes
\begin{eqnarray}
    z_{\rm cor} = f_{\rm gain} (z_{\rm fit}+1)-1,
    \label{eq:zcor}
\end{eqnarray}
where $z_{\rm cor}$ and $z_{\rm fit}$ are the corrected redshift and
that obtained from the fitting results, respectively.

Although the above-mentioned correction is obviously valid for the regions
near the calibration sources, the gain could depend upon the position on the same CCD
chip owing to the charge transfer inefficiency (CTI). As for this, an accuracy of 0.2\% was reported
over the CCD chips before spaced-row charge injection (SCI) was adopted \citep{Ota07}.
Matsumoto et al. (2007)\footnote{JX-ISAS-SUZAKU-MEMO-2007-06} reported that CTI was almost zero 
just after the SCI was adopted. Since our observation was held at the timing very near to this report,
we conclude that the systematic errors of the energy scale because of CTI are 
estimated to be 0.2\%, which corresponds to be $\sim 600$ km s$^{-1}$.

\begin{figure}
  \begin{center}
    \FigureFile(80mm,80mm){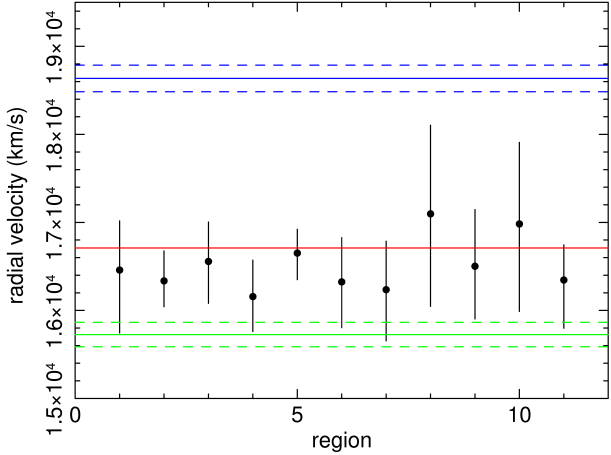}
  \end{center}
  \caption{Line-of-sight velocities of the ICM for the regions presented in figure 
           \ref{fig8}. Red , green, and blue solid lines show the mean 
            line-of-sight velocities of the member galaxies for the entire A2319, 
            A2319A subgroup, and A2319B subgroup, respectively. 
            The error bars stand for only statistical ones.
           }\label{fig9}
\end{figure}

We fit spectra of each XIS sensor with APEC model in the energy band
of 5.0 -- 10.0 keV. The resultant values of $z_{\rm fit}$ are corrected
into $z_{\rm cor}$ in a way of equation (\ref{eq:zcor}), which are
converted into line-of-sight velocities. Then, we calculate arithmetic means of the results
of all four XIS sensors.
Figure \ref{fig9} shows line-of-sight velocities of the ICM for the
regions presented in figure \ref{fig8}, where  red , green, and blue
solid lines represent the mean line-of-sight velocities of the member
galaxies for the entire A2319, A2319A subgroup, and A2319B subgroup,
respectively. The error bars stand for only statistical ones.
Basically, the obtained ICM velocities lie between those of
entire A2319 and A2319A. No significant velocity difference is detected within
the observed region. 
Taking account of the systematic errors because of CTI, we confirm that the observed ICM velocities 
are consistent with those of entire A2319 and A2319A subgroup, 
and do not find any signs of ICM motion associated with A2319B subgroup in the observed region.

\begin{figure}
  \begin{center}
    \FigureFile(80mm,80mm){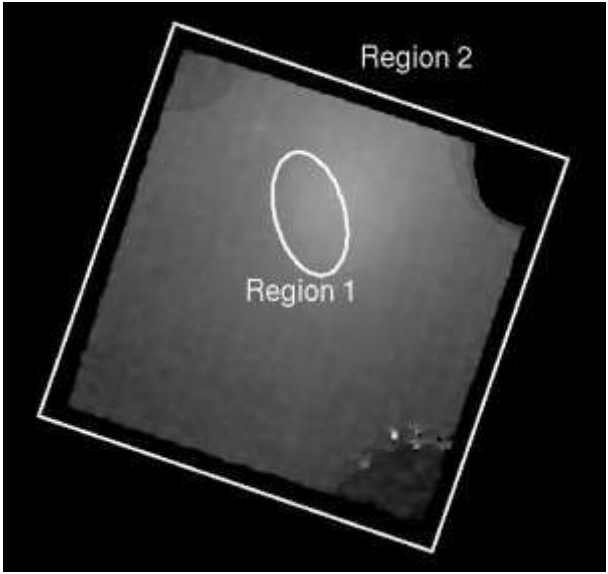}
  \end{center}
  \caption{Regions used in measurement of the line-of-sight velocity around the cold front
           in figure \ref{fig11}.
           }\label{fig10}
\end{figure}

\begin{figure}
  \begin{center}
    \FigureFile(80mm,80mm){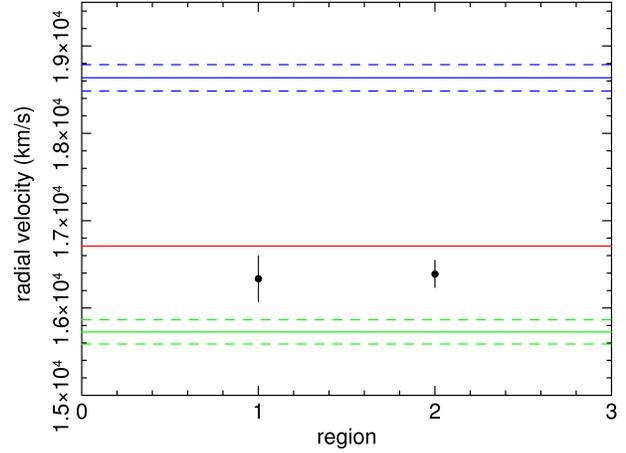}
  \end{center}
  \caption{Same as figure \ref{fig9}, but for the regions presented in figure \ref{fig10}.
           }\label{fig11}
\end{figure}

As mentioned before, a cold front is found in the south-east of the
X-ray center by Chandra observations \citep{Ohar04, Govo04}. In addition, we find
that the region in the north of the cold front has lower temperature and
higher abundance. This may mean that there exits a cold core originated
from the subcluster that infalls in the past. Therefore, it is possible
that ICM in this region has negligible relative velocity to the surrounding ICM.
In order to examine this, we analyze spectra of the regions presented in
figure \ref{fig10} and measure the line-of-sight velocities in a similar way.
The results are shown in figure \ref{fig11}. Again, no significant
velocity difference is detected. The obtained ICM velocities lie between those of
entire A2319 and A2319A. 
It is also clear that there are no signs of ICM motion related with A2319B subgroup.
The size of Region 1 in figure \ref{fig10} is approximately $3'\times5.5'$.
Taking account of the X-ray telescope response, we estimate that
$\sim$ 30\% of photons detected in the Region 1 are originated from Region 2,
and less than 10\%, vice versa, respectively.
Therefore, this contamination does not affect our results significantly.

\subsection{Hard X-ray Properties and Constraint of Non-thermal Components}
\begin{figure}
  \begin{center}
    \FigureFile(80mm,80mm){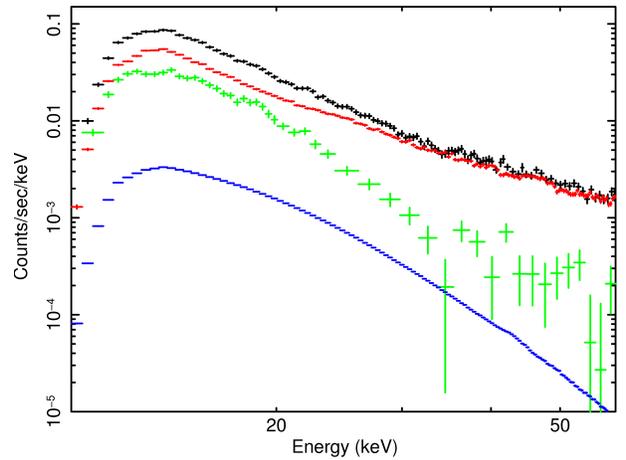}
  \end{center}
  \caption{The PIN spectrum of this observation. Black, red, and green crosses show the
           spectrum of the data, NXB model, and residual signals (data-NXB),
           respectively. A typical CXB model spectrum is also presented in blue crosses.
           }\label{fig12}
\end{figure}

The PIN spectrum of this observation is shown in Figure \ref{fig12}, where
black, red, and green crosses represent the data, NXB model, and
residual signals (data-NXB), respectively. A typical CXB model spectrum is
also shown in blue crosses. Hard X-ray emission from A2319 is clearly
detected in the energy below $\sim$ 40 keV. Therefore, we use the PIN spectrum 
in the energy band of 13.0-40.0 keV hereafter. 
We also check the GSO spectrum
and confirm that the data are consistent with NXB. Thus, we do not use them.

First, we fit the PIN spectrum alone with the APEC or power-law model in the energy range of 
13.0-40.0 keV. The metal abundance and redshift in the APEC model are fixed to be 0.3 and 0.0557, 
respectively, because it is very difficult to determine these parameters with only PIN data. 
The results of the fit are presented in table \ref{tab:pin_spec}. 
It is obvious that the spectrum is better represented by a thermal APEC model 
than a non-thermal power-law one. The temperature determined from the PIN spectrum
is similar to the values in figure \ref{fig4} determined from the XIS data.
\begin{table}
  \begin{center}
    \caption{Fitting results of the PIN spectrum}
    \label{tab:pin_spec}
    \begin{tabular}{ccccc}
      \hline\hline
      model   &   $\Gamma$ or $kT$(keV) & $\chi ^2$/d.o.f   \\
      \hline
      $PL$    &   $3.1^{+0.1}_{-0.1}$    &  81.7/69          \\
      $1kT$   &   $10.9^{+0.9}_{-0.9}$   &  67.7/69          \\
      \hline
    \end{tabular}
  \end{center}
\end{table}

Second, we perform joint spectral analysis of PIN and XIS to investigate hard X-ray properties.
It should be noted that soft X-ray spectral information is essential to constrain 
spectral components characteristic in the hard X-ray band such as non-thermal power-law and/or very
hot thermal ones. We use the XIS spectrum in the energy band of 2.0-10.0 keV and 2.0-8.0 keV
for FI and BI CCDs, respectively. FI CCD spectra and associated response files 
are summed, and FI, BI, and PIN spectra are fitted simultaneously. 
Systematic errors between XIS and PIN normalization are taken into account properly.
We tried to fit XIS+PIN spectra in a similar way but for using XIS spectra also below 2 keV.
The fitting results are not acceptable if we adopt common $N_{H}$ for FI and BI CCDs,
because significant residuals inconsistent between FI and BI CCDs are found in the soft band.
Then, we tried to fit the spectra for each XIS sensor and PIN simultaneously allowing
that each XIS has a different $N_{H}$ value. This improves the fitting results significantly.
We suspect that this is possibly because the contamination correction of arf files are not very good 
in case of regions as large as almost entire XIS FOV, 
though we have no definitive idea about this problem at present. 
In addition, it is possible that this kind of systematic errors become apparent
because the spectra have sufficiently good statistics.
The difference of $N_{H}$ among four XIS sensors is typically $\sim 1-3 \times 10^{20}$cm$^{-2}$, 
which is so small that it does not seriously affect the results of the spectral fit above 2.0 keV.
Thus, we decide not to use XIS spectra below 2.0 keV to avoid this unsolved issue.

Since we are interested in both non-thermal inverse Compton and very hot components, 
spectral models used here are a single temperature APEC model ($1kT$), single temperature  APEC plus
power-law model ($1kT+PL$), two temperature APEC model ($2kT$), and two temperature APEC plus power-law 
model ($2kT+PL$). Again, we fix both $N_H=7.93 \times 10^{20}$ cm$^{-2}$ and redshift=0.0557 as in 
\S\ref{ss:tem_ab}. In the $2kT$ and $2kT+PL$ models, 
a relative normalization ratio between two APEC models is 
allowed not to be common between XIS and PIN, to compensate small ($\sim10$\%) uncertainty 
in the arf generation.
It is very difficult to constrain power-law components
in the XIS energy range because of dominant presence of thermal emission, which could degrade 
the HXD fitting results. In order to avoid this, thus, a relative normalization ratio between APEC 
and power-law models is also allowed not to be common between XIS and PIN.
A photon index of the power-law component is also fixed assuming that
it is emitted via inverse Compton process of CMB from the same electron population that radiates
synchrotron radio. Considering that typical magnetic field strength of the intracluster space is
$\sim \mu$G, electrons that radiates the inverse Compton hard X-ray of 20-40 keV emit relatively low
frequency synchrotron radio. We adopt 1.92 or 2.4 as the photon index of the power-law component
because 0.92 (408-610MHz) and 1.4 (26-610MHz) are reported as radio spectral index
\citep{Eric78, Harr78, Fere97}.

\begin{figure}
  \begin{center}
    \FigureFile(80mm,80mm){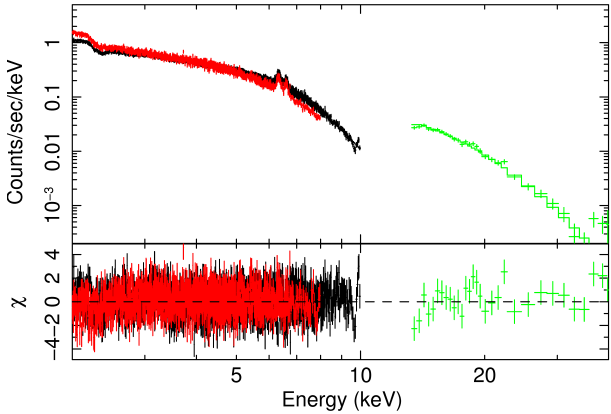}
  \end{center}
  \caption{The wide band spectrum of the A2319 fitted with the $1kT$ model. Black, red, and green 
           crosses show the spectrum of XIS-FI, XIS-BI, and HXD-PIN, respectively. 
           The best fit model is presented in dashed histograms.
           }\label{fig13}
\end{figure}

\begin{figure}
  \begin{center}
    \FigureFile(80mm,80mm){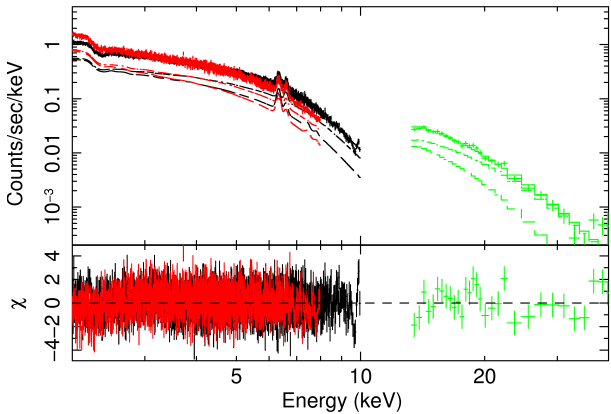}
  \end{center}
  \caption{Same as figure \ref{fig13}, but for the $2kT$ model. 
           Two thermal components are independently presented as dashed histograms.
           }\label{fig14}
\end{figure}

\begin{figure}
  \begin{center}
    \FigureFile(80mm,80mm){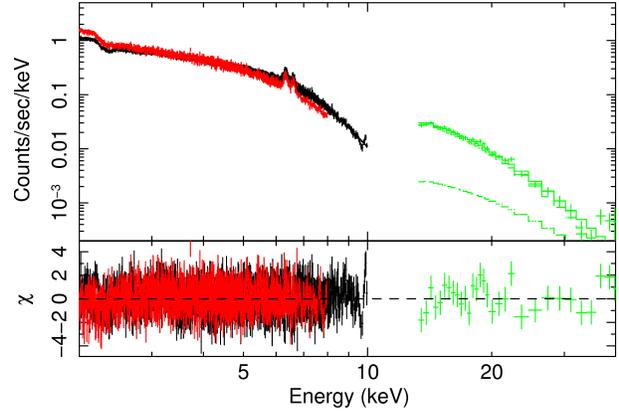}
  \end{center}
  \caption{Same as figure \ref{fig13}, but for the $1kT+PL$ model with a photon index 1.92.
           Thermal and power-law components are independently presented as dashed histograms for
           HXD. Only thermal component is presented for XIS.
           }\label{fig15}
\end{figure}

Figure \ref{fig13}, \ref{fig14}, and \ref{fig15} show the spectral fit results for
the $1kT$, $2kT$, and $1kT+PL$ model with the photon index 1.92, respectively.
Best fit parameters of the $1kT$ and $2kT$ models are listed in table \ref{tab:1kT2kT}.
Although the reduced chi square is slightly improved in the $2kT$ model,
the $1kT$ model gives us sufficiently acceptable results. Thus, it is possible that 
there exits a little very hot component ($\sim 16$keV), which does not have to be needed.
The detailed fitting results of the $1kT+PL$ and $2kT+PL$ models are also presented in table 
\ref{tab:1kT+PL} and \ref{tab:2kT+PL}, respectively.
Clearly, adding a power-law component does not improve the fitting results significantly 
in both the $1kT$ and $2kT$ models. 
\begin{table}
  \caption{Best fit parameters of the $1kT$ and $2kT$ models for XIS and PIN spectra.}
  \label{tab:1kT2kT}
  \begin{center}
    \begin{tabular}{lll}
      \hline \hline
                               &   $1kT$                &   $2kT$               \\
      \hline
         $kT_{\rm low}$(keV)    &  $9.7^{+0.1}_{-0.1}$    &  $6.5^{+0.5}_{-0.5}$    \\
         $kT_{\rm high}$(keV)   &     $-$                &  $15.7^{+1.2}_{-1.0}$   \\
         $Z$($Z_{\odot}$)       &  $0.26^{+0.01}_{-0.01}$ &  $0.28^{+0.01}_{-0.01}$  \\
         $N_{\rm low, XIS}$      &  $0.17^{+0.00}_{-0.00}$ &  $8.5^{+1.1}_{-1.1} \times 10^{-2}$ \\
         $N_{\rm low, PIN}$      &  $0.18^{+0.01}_{-0.01}$ &  $0.14^{+0.04}_{-0.04}$   \\
         $N_{\rm high, XIS}$     &     $-$                &  $9.2^{+1.1}_{-1.2} \times 10^{-2}$  \\
         $N_{\rm high, PIN}$     &     $-$                &  $6.4^{+1.4}_{-1.3} \times 10^{-2}$  \\
         $\chi^2/{\rm d.o.f.}$ &  3215.6/2978           &   3136.6/2975            \\
      \hline
    \end{tabular}
  \end{center}
\end{table}
\begin{table}
  \caption{Same as table \ref{tab:1kT2kT}, but for the $1kT+PL$ models, 
           where a photon index of the power low component is fixed to be 1.92 or 2.4.}
  \label{tab:1kT+PL}
  \begin{center}
    \begin{tabular}{lll}
      \hline \hline
                               &   $1kT+PL(1.92)$                &   $1kT+PL(2.4)$      \\
      \hline
         $kT_{\rm low}$(keV)    &  $9.7^{+0.1}_{-0.1}$    &  $9.7^{+0.1}_{-0.1}$    \\
         $Z$($Z_{\odot}$)       &  $0.26^{+0.01}_{-0.01}$ &  $0.26^{+0.01}_{-0.01}$  \\
         $N_{\rm low, XIS}$      &  $0.17^{+0.00}_{-0.00}$ &  $0.17^{+0.00}_{-0.00}$  \\
         $N_{\rm low, PIN}$      &  $0.16^{+0.02}_{-0.01}$ &  $0.15^{+0.02}_{-0.02}$   \\
         $\Gamma_{\rm PL}$      &   1.92(fixed)          &  2.4(fixed)             \\
         $N_{\rm PL, XIS}$      &   $0.0^{+0.7}_{-0.0} \times 10^{-3}$  & $0.0^{+0.3}_{-0.0} \times 10^{-3}$ \\
         $N_{\rm PL, PIN}$      &  $3.8^{+2.8}_{-2.8} \times 10^{-3}$   & $2.2^{+1.8}_{-1.7} \times 10^{-2}$ \\
         $\chi^2/{\rm d.o.f.}$ &  3210.9/2976           &   3211.2/2976            \\
      \hline
    \end{tabular}
  \end{center}
\end{table}
\begin{table}
  \caption{Same as table \ref{tab:1kT+PL}, but for the $2kT+PL$ models.}
  \label{tab:2kT+PL}
  \begin{center}
    \begin{tabular}{lll}
      \hline \hline
                               &   $2kT+PL(1.92)$                &   $2kT+PL(2.4)$      \\
      \hline
         $kT_{\rm low}$(keV)    &  $6.5^{+0.5}_{-0.6}$    &  $6.5^{+0.5}_{-0.6}$    \\
         $kT_{\rm high}$(keV)   &  $15.7^{+0.9}_{-1.1}$   &  $15.7^{+0.9}_{-1.1}$   \\
         $Z$($Z_{\odot}$)       &  $0.28^{+0.01}_{-0.01}$ &  $0.28^{+0.01}_{-0.01}$  \\
         $N_{\rm low, XIS}$      &  $8.6^{+1.1}_{-1.1} \times 10^{-2}$ &  $8.6^{+1.1}_{-1.1} \times 10^{-2}$ \\
         $N_{\rm low, PIN}$      &  $0.15^{+0.04}_{-0.04}$ &  $0.15^{+0.04}_{-0.04}$   \\
         $N_{\rm high, XIS}$     &  $9.2^{+1.2}_{-1.2} \times 10^{-2}$  &  $9.2^{+1.2}_{-1.2} \times 10^{-2}$ \\
         $N_{\rm high, PIN}$     &  $6.3^{+1.5}_{-2.6} \times 10^{-2}$  &  $6.3^{+1.5}_{-2.9} \times 10^{-2}$ \\
         $\Gamma_{\rm PL}$      &  1.92(fixed)          &   2.4(fixed)             \\
         $N_{\rm PL, XIS}$      &   $0.0^{+0.4}_{-0.0} \times 10^{-3}$   &  $0.0^{+3.2}_{-0.0} \times 10^{-4}$ \\
         $N_{\rm PL, PIN}$      &   $0.0^{+1.3}_{-0.0} \times 10^{-2}$   &  $0.0^{+3.3}_{-0.0} \times 10^{-2}$ \\
         $\chi^2/{\rm d.o.f.}$ &  3136.7/2973           &   3136.7/2973            \\
      \hline
    \end{tabular}
  \end{center}
\end{table}

Our spectral modeling for thermal components are somewhat too simple considering
that Chandra observations \citep{Ohar04, Govo04} show multi temperature structures with a certain 
temperature range. In general, if multi-temperature spectra are forced to be fitted 
with a simple $1kT+PL$ model, the power-law component tends to be overestimated. 
Thus, the derived upper limit of the power-law component can be regard as relatively conservative
one. In case of the $2kT$ modeling, on the other hand, the situation could be more complicated.
The higher energy part of the spectra can be represented by both higher temperature and/or
power-law components. It is not trivial how this increase of the degree of the freedom of the model
affects the actual fitting results.

In order to constrain an upper limit of a non-thermal power-law component, not only statistical errors 
but also systematic ones of both CXB and NXB have to be taken into account properly. It is well-known
that the CXB fluctuations can be modeled as 
$\sigma_{\rm CXB}/I_{\rm CXB} \propto \Omega_e^{-0.5} S_c^{0.25}$, 
where $\Omega_e$ and $S_c$ are the effective solid angle and upper cutoff flux of a point source, 
respectively. From the HEAO-1 A2 results, $\sigma_{\rm CXB}/I_{\rm CXB}=2.8$\% 
with $\Omega_e=15.8$ deg$^2$ and $S_c = 8 \times 10^{-11}$ erg s$^{-1}$ cm$^{-2}$ \citep{Shaf83}. 
Considering a rough estimate of NXB uncertainties of PIN 
($\sim 5 \times 10^{-12}$ erg s$^{-1}$ cm$^{-2}$ ), we adopt conservative upper cutoff
flux $S_c \sim 8 \times 10^{-12}$ erg s$^{-1}$ cm$^{-2}$. 
Thus, the CXB fluctuation in the HXD PIN FOV is expected
to be 18\% at the 90\% confidence level because of $\Omega_e =0.32$ deg$^2$. 
\citet{Fuka09} reports that the reproducibility of blank sky observations 
separated into 10 ks exposures give distribution of 5.7\% at the 90\% confidence level, 
where contribution from the statistical error 
of typically 3.3 \% or larger and the effect of CXB fluctuation (1.3\% of the total background)
are included. This means that the NXB systematic uncertainty is 4.5\% 
($=\sqrt{(5.7)^2-(3.3)^2-(1.3)^2}$)
at the 90\% confidence level.

In the joint spectrum analysis of PIN and XIS, NXB and CXB components are fluctuated at the 90 \%
confidence level of the systematic uncertainty mentioned-above 
(4.5\% and 18\% for NXB and CXB, respectively)
. This causes the changes of 
the best-fit parameters in the fits and gives us the systematic errors.
Including both the statistical errors and systematic ones of CXB and NXB, we derive 
an upper limit of a power-law component in 10-40 keV at the 90 \% confidence level for each model.
The results are presented in table \ref{tab:uplimfl}, where the former and latter errors of flux are
statistical and systematic ones, respectively.
Although the results slightly depend on spectral modeling, all models give us the upper limit of 
$\sim 3 \times 10^{-11}$ erg s$^{-1}$ cm$^{-2}$. 
\begin{table}
  \caption{Flux of a power-law component in 10-40 keV with statistical and systematic errors and 
           its upper limit at the 90 \% confidence level for each model.}
  \label{tab:uplimfl}
  \begin{center}
    \begin{tabular}{lll}
      \hline \hline
            model      &   flux                     &   upper limit        \\
                       & (erg s$^{-1}$ cm$^{-2}$)  & (erg s$^{-1}$ cm$^{-2}$) \\  
      \hline
        $1kT+PL(1.92)$ &  $1.1^{+0.8+1.3}_{-0.8-1.1} \times 10^{-11}$  & $<2.6 \times 10^{-11}$    \\
        $1kT+PL(2.4)$  &  $1.5^{+1.2+1.9}_{-1.2-2.1} \times 10^{-11}$  & $<3.8 \times 10^{-11}$    \\
        $2kT+PL(1.92)$ &  $0.0^{+3.7+2.2}_{-0.0-0.0} \times 10^{-11}$  & $<4.3 \times 10^{-11}$    \\
        $2kT+PL(2.4)$  &  $0.0^{+2.2+3.9}_{-0.0-0.0} \times 10^{-11}$  & $<4.5 \times 10^{-11}$    \\        
      \hline
    \end{tabular}
  \end{center}
\end{table}

\section{Discussion}

\subsection{Temperature and Abundance Structures}
Complex temperature structures are obtained in the central region of
the Abell 2319 by Chandra observations \citep{Ohar04, Govo04}. Basically,
our results are consistent with theirs though FOV of both observations
are not perfectly overlapped with each other. There are high temperature
regions in the north east of the X-ray center (region 1 and 7 in figure
\ref{fig3}), and regions in the north-west of the cold front (region 2, 3, and 4 in figure
\ref{fig3}) have lower temperature. However, very high temperature
($\sim 15$ keV) components in the Chandra data are not found in our
2D spectral analysis of the XIS data alone.
On the other hand, the $2kT$ model fitted to the XIS+PIN spectrum shows a strong 15 keV
component, but this is likely an artifact from introducing a strong 6.5 keV
component for a cluster with average temperature of 9.7 keV.
Since the spectral fit with only HXD prefers a temperature less than 11.8 keV,
and the XIS+PIN wide band spectrum is well fitted by a 9.7 keV single temperature
thermal model,
we conclude that there is no strong evidence of a bright very hot component,
as high as 15 keV in the Suzaku data.
It should be noted that temperature determination of very hot
gas with temperature more than 10 keV is difficult for Chandra CCD without any sensitivity above
10 keV. It is also possible that our 2D XIS results are significantly spatially
smoothed because of the moderate spatial resolution of Suzaku.

\citet{Mark96} studied a larger scale temperature structure of the A2319 with
ASCA. It reveals that the A2319b region and outer parts of the cluster
have lower temperature, which are without FOV of XIS but partly within
that of PIN. Our wide band spectral analysis of the $2kT$ model shows 
that a relative ratio of normalization between two APEC models in XIS and PIN
are significantly different than each other. The results tell us that 
the lower temperature gas is more dominant in the PIN spectrum than XIS,
which is not surprising and naturally expected from the above-mentioned ASCA results.

Our 2D and projected radial spectrum analysis show that metal abundance in the central
cool region is clearly higher than the surrounding region (see figure
\ref{fig5} and \ref{fig7}). \citet{Mole99} performed similar analysis
with Beppo-SAX, and shows that the metal abundance distribution is consistent with homogeneous one
within statistical errors.
Because of large effective area and high sensitivity of XIS,  metal
abundance inhomogeneity in A2319 is clearly detected for the first time with Suzaku.

\subsection{Bulk Flow Motion in the ICM}
As written before, though no significant line-of-sight velocity difference
within the observed region is detected, we find that their values are
consistent with those of entire A2319 and A2319A subgroup for the first time.
In figure \ref{fig9}, the largest velocity difference is found between
region 4 and 8, which provides us with the value $940^{+1083}_{-1131}$
km s$^{-1}$. The sound velocity of 10 keV ICM is
$\sim 1700$ km s$^{-1}$. Thus, it is probable that the internal motion
of the ICM in the observed region is subsonic, though possibly it is close
to the sound velocity. 
On the other hand, the difference between
the observed region and A2319b is nearly 3000 km s$^{-1}$. This means that 
super sonic ICM collision is expected if the ICM associated with the A2319b has
the velocity similar with the A2319b member galaxies. Another possibility
is that the A2319b group is not gravitationally bound to the A2319 but
located along the line-of-sight by chance. Please note that this possibility is not negligible
from orbital motion analysis with a two-body model in \citet{Oege95}.

\subsection{Constraint of the Nonthermal Emission and Magnetic Field Strength}
Let us constrain the magnetic field strength from the combined analysis of radio synchrotron
flux and an inverse Compton hard X-ray upper limit \citep{Rybi79}. 
Typical energy $h \nu'$ of photons scattered via
inverse Compton processes by electrons with energy $\gamma m_{\rm e} c^2$ is 
$h \nu' \simeq 4 \gamma^2 h \nu /3$, where $h \nu$ is the photon energy before the scattering.
With typical CMB photon energy ($h \nu \simeq 2.3 \times 10^{-4}$ eV), the range of electron's 
relativistic gamma factor corresponding to the 10-40 keV inverse Compton hard X-ray becomes 
$3.3 \times 10^7 < \gamma^2 < 1.3 \times 10^8$. 
On the other hand, a typical synchrotron radio frequency (or, a critical frequency) emitted
by the electrons with the magnetic field strength $B$ is 
$(\nu_c / {\rm MHz}) \simeq 3.3 \gamma^2 (B/{\rm G})$,
where homogeneous pitch angle distribution is assumed.
Therefore, typical synchrotron radio frequency range emitted by the electrons in the 
above-mentioned energy range becomes
$1.1 \times 10^8 (B/{\rm G}) < (\nu_c / {\rm MHz}) < 4.3 \times 10^8 (B/{\rm G})$.
It is reported that radio flux of A2319 halo is 1.0 Jy at 610MHz and that its spectral index is
0.92 (408-610MHz) or 1.4 (26-610MHz) \citep{Harr78,Fere97}. 
Thus, radio flux corresponding 
to the energy range of 10-40 keV inverse Compton hard X-ray becomes values 
as in table \ref{tab:radioflux}
for each spectral index, 
with a monochromatic approximation for a single electron's spectrum and 
on the assumption 
of a single power-law electron energy distribution.

\begin{table}
  \caption{Radio flux in the energy band corresponding to the 10-40 keV inverse Compton component}
  \label{tab:radioflux}
  \begin{center}
    \begin{tabular}{lll}
      \hline \hline
       photon index    &   radio flux ($F_{\rm sync}$)  \\
                       & (erg s$^{-1}$ cm$^{-2}$)       \\  
      \hline
        $1.92$         &  $2.3 \times 10^{-13} (B/{\rm G})^{-0.08}$  \\
        $2.4$          &  $5.1 \times 10^{-16} (B/{\rm G})^{-0.4}$   \\
      \hline
    \end{tabular}
  \end{center}
\end{table}

Flux of the inverse Compton scattering of CMB photons and synchrotron radiation 
from the same electron population has the following relationship,
\begin{eqnarray}
   \frac{F_{\rm IC}}{F_{\rm syn}} =\frac{U_{\rm CMB}}{U_{\rm mag}}=\frac{U_{\rm CMB}}{B^2/8\pi},
\end{eqnarray}
where $U_{\rm CMB}$ and $U_{\rm mag}$ are the energy density of CMB photons and magnetic field, 
respectively.
With $U_{\rm CMB}=5.2 \times 10^{-13}$ erg cm$^{-3}$ at $z=0.0557$ and the values 
in table \ref{tab:uplimfl} and \ref{tab:radioflux}, the lower limit of the magnetic field strength
for each model is obtained as in table \ref{tab:lowlimmag}.
Although the detailed results slightly depend on the spectral modeling, basically the field strength
tends to be  more than $\sim 0.2$ $\mu$G.
\begin{table}
  \caption{The lower limit of the magnetic field strength for each model.}
  \label{tab:lowlimmag}
  \begin{center}
    \begin{tabular}{lll}
      \hline \hline
            model      &  $B$($\mu$G) \\
      \hline
        $1kT+PL(1.92)$ &  $>0.19$     \\
        $1kT+PL(2.4)$  &  $>0.27$     \\
        $2kT+PL(1.92)$ &  $>0.14$     \\
        $2kT+PL(2.4)$  &  $>0.25$     \\        
      \hline
    \end{tabular}
  \end{center}
\end{table}

A2319 was observed by Beppo-SAX PDS in the hard X-ray band \citep{Mole99}. 
The obtained upper limit flux of the power-law component in the energy band of 13-50 keV is 
2.3 $\times 10^{-11}$ or 2.0 $\times 10^{-11}$ erg s$^{-1}$ cm$^{-2}$, 
depending on the adopted radio spectral index 
0.92 (408-610 MHz) or 2.2 (610-1420 MHz), respectively. Different radio spectral indexes are used 
because the PDS results rely on higher energy band than PIN. 
When these results are compared with ours, however, it should be noted that the Crab's 
20-80 keV flux derived from Beppo-SAX PDS is 21 \% smaller than that derived from Suzaku. Thus,
the above-mentioned Beppo-SAX results can be converted into Suzaku-equivalent fluxes of 
2.9 $\times 10^{-11}$ or 2.5 $\times 10^{-11}$ erg s$^{-1}$ cm$^{-2}$ 
for the radio spectral index 0.92 or 2.2, respectively.
These results seem to be similar to our results of the $1kT+PL(1.92)$ model at first glance.
However, please remember that HXD PIN FOV is much narrower than Beppo-SAX PDS,
which is preferable in order to avoid contamination sources, 
but large enough to cover the A2319 radio halo entirely. 

Recently, \citet{Ajel09} studied the hard X-ray properties of 10 galaxy clusters including 
Abell 2319 using Swift-BAT. The obtained upper limit of the power-law component in the energy band 
of 10-40 keV is 2.9 $\times 10^{-12}$ or 1.7 $\times 10^{-12}$ erg s$^{-1}$ cm$^{-2}$ with a radio 
spectral index 0.92 through Swift-BAT alone or Swift-BAT + XMM-Newton analysis, respectively.
Here, their original values at the $3\sigma$ confidence level in the energy band of 50-100 keV
are converted into those at the 90\% confidence level in the 10-40 keV band.
Considering that A2319 has relatively high temperature ($\sim 10$ keV) and that
Swift-BAT, with its long exposure, has better sensitivity in the higher energy band ($\sim 100$ keV) 
than Suzaku PIN, it is not 
so surprising that their results give tighter constraint. Suzaku is superior in determining the thermal 
properties, thanks to its high sensitivity around 20-40 keV.
We note that their power-law "upper-limit" is quite low, much less than those inferred from
the sensitivity. Their sensitivity plot shows $\sim$ 1 mCrab, or $\sim 1.6 \times 10^{-11}$ at 10-40 keV, 
as a 3-sigma sensitivity, or even worse above 100 keV. Thus, Swift-BAT is presenting $\sim$ 2 or less 
$\times 10^{-12}$ erg s$^{-1}$ cm$^{-2}$ upper-limit using the detector with a 90\% confidence sensitivity 
of $\sim$ 7 $\times 10^{-12}$ erg s$^{-1}$ cm$^{-2}$. This is of course allowed in some cases, 
especially when the photon statistics favors negative flux. Anyway, our results, independently obtained, 
are consistent with theirs. We would also like to point out that the Swift alone analysis
lacks soft X-ray spectral information important to determine thermal component precisely,
and that generally speaking cross-calibration between detectors onboard different satellites 
(such as XMM and BAT) is not so easy.

As we wrote before, ASCA results \citep{Mark96} suggest there is lower temperature ($\sim 8$ keV) ICM 
in the outer region without XIS but within PIN FOV, though this is not clear in Beppo-SAX results
\citep{Mole99}.
The existence of this cold ICM could affect the estimation of the upper limit of the
power-law component. To evaluate this, we make the spectra assuming that the outer region ($r>10'$)
has 8 keV and perform joint spectrum analysis of XIS and PIN in a similar way in subsection 3.3.
The resultant flux of the power-law component in 10-40 keV increases 
by $10.4 \times 10^{-12}$ and $9.9 \times 10^{-12}$ erg s$^{-1}$ cm$^{-2}$ 
for the $1kT + PL$ models with a photon index 1.92 and 2.4, respectively.
These values are smaller than both the statistical errors and those
induced by background uncertainty,
as listed in table 6. This effect shall affect the Beppo-SAX PDS
results as well.

\subsection{The Energy Budget of the Intracluster Space}
From the observed results, we can get information about the energy densities of the thermal ICM, 
non-thermal electrons, and magnetic field in the intracluster space of the A2319, which are basic and
important parameters for the cluster structure and evolution. For simplicity, 
we assume that the radio halo region is a sphere with a radius 0.66 Mpc. 
With $\beta$-model parameters obtained by ROSAT
\citep{Trev00} and $kT=10$ keV, energy density of the thermal ICM becomes 
$U_{\rm th} = 4 \time 10 $ eV cm$^{-3}$.
From the upper limit of an inverse Compton component of the $1kT+PL(1.92)$ model 
in table \ref{tab:uplimfl}, we obtain the energy density of the relativistic electrons 
corresponding to 10-40 keV hard X-ray 
(or, $5.7 \times 10^3 < \gamma < 1.1 \times 10^4$) $U_e < 2 \times 10^{-2}$ eV cm$^{-3}$.
Therefore, $U_e/U_{\rm th} < 5 \times 10^{-4}$, though in this calculation
we do not consider the contribution from relatively low energy electrons, which could be dominant
in the energy density of the non-thermal electron populations. 
With the lower limit of the magnetic field for 
the $1kT+PL(1.92)$ model in table \ref{tab:lowlimmag}, the energy density of the magnetic field
becomes $U_{\rm mag} > 1 \times 10^{-3}$ eV cm$^{-3}$, 
which means $U_{\rm mag}/U_{\rm th} > 3 \times 10^{-5}$.

\subsection{A Particle Acceleration Scenario}
In this subsection, let us discuss a particle acceleration scenario of the A2319
implied by the XIS and HXD results mentioned above. 
The XIS results tell us two important points as follows.
First, the central cool region associated with the cold front have higher metal abundance, 
which suggests that this might be an old remnant of a subcluster's cool core that infalls
in the past. Second, ICM velocity difference in the observed region is probably subsonic, which means 
that no definitive shocks are expected there. This is also supported by the fact that there is not 
a significant very hot component from the XIS+HXD analysis. 
On the other hand, existence of subsonic turbulence ($\Delta v \sim 500$ km s$^{-1}$) can be allowed. 
Therefore, it is more likely that non-thermal electrons relevant to 
the radio halo of the A2319 are accelerated 
by the intracluster turbulence rather than the shocks.
The turbulence is probably excited by the past merger activity, which is also responsible for 
the cold front and central cool core with higher abundance. This picture is also consistent with 
the fact that numerical simulations show that the 
turbulence motion is developed in the late phase of mergers \citep{Taki05}.
The discussion on the energy budget in the previous subsection shows that 
the magnetic field energy density is likely much less than the hydrodynamical turbulence,
which means that the magnetic turbulence can be easily excited \citep{Roet99, Asai07, Taki08}.
Although a collision between the subgroup A2319A and A2319B is likely supersonic, 
it does not seem to form sufficiently strong shocks to excite the radio halo activity. 
It is probable that they are not so close to each other that sufficiently strong shocks occur.

\subsection{Future Prospects}

We obtained only an upper limit of the ICM line-of-sight velocity difference. 
To resolve the intracluster turbulence ($\Delta v \sim 500$ km s$^{-1}$), 
a different type of X-ray spectrometer than CCD is highly desired.
The ASTRO-H satellite \citep{Taka06}, which is planned to be launched around 2013, will enable us 
to measure the line-of-sight velocity directly with the X-ray microcalorimeters and 
provide us with useful information on the dynamical status of the ICM. ASTRO-H is also planed to have
hard X-ray mirrors and imager, which enable us to get hard X-ray images at very high sensitivity.
This gives us very important clues to understand where and how the particle acceleration processes
occur in the intracluster space.

\section{Conclusions}
We observed the central region of the Abell 2319 with Suzaku. From the XIS analysis, 
we found that the central cool region in the north-west of the cold front 
has higher metal abundance than the surrounding region for the first time.
We measure the line-of-sight velocities of the ICM with XIS and find that the velocity difference
is less than $\sim$ 2000 km s$^{-1}$, which means that the ICM motion is probably subsonic.
In addition, we show that the velocities in the observed region are consistent with those of
entire A2319 and A2319A subgroup for the first time. 
No signs of ICM motion related to A2319B subgroup are found.
If the ICM associated with the A2319B subgroup has the similar velocity of A2319B member galaxies, 
a supersonic collision is expected in the ICM. 
From the XIS+HXD wide band spectral analysis, we search for very hot and
non-thermal inverse Compton components. 
There can be a little amount of very hot ($\sim$ 15 keV) component,
but it does not have to exist. No significant inverse Compton component is detected, and we derive
its upper limit for several spectral models. Typically, it becomes 
$< 3 \times 10^{-11}$ erg s$^{-1}$ cm$^{-2}$, which is consistent with the recent Swift-BAT results. 
Comparing the synchrotron radio flux and upper limit 
of the inverse Compton one, we obtain the lower limit of the magnetic field strength 
$B > 0.2 \mu$G, though the results slightly depend upon the spectral modeling. 
Taking into account of the lack of
the supersonic ICM motion and a very hot component, we conclude that the relativistic electrons 
responsible for the radio halo of the Abell 2319 are more likely 
accelerated by the turbulence rather than shocks.

\bigskip
The authors would like to thank K. Matsushita and S. Shibata for helpful comments,
and an anonymous referee for his/her useful comments and constructive suggestions.
We are also grateful to the Suzaku operations team for their support
in planning and executing this observation.
M. T. was supported in part by a Grant-in-Aid from the
Ministry of Education, Science, Sports, and Culture of Japan (19740096).


\end{document}